\documentclass[a4paper]{nature}

\pdfoutput=1
\usepackage[T1]{fontenc}
\usepackage{ifthen,relsize,multicol,marvosym,float}
\usepackage{tabularx,tabulary,multirow,lastpage,lineno}
\usepackage[pdftex]{graphicx}
\usepackage{epic,eepic,epsfig}
\usepackage{siunitx}
\usepackage[polutonikogreek,french,german,english]{babel}
\usepackage{caption}

\usepackage{titling}
\usepackage[centertags]{amsmath}
\usepackage{pbox,fancyhdr,textcomp}
\usepackage{amssymb,amsfonts,oldgerm,mathrsfs,ipa}
\usepackage[text={6in,8.25in},centering]{geometry}

\newcommand{\afourlengths}{\geometry{a4paper,text={15cm,22.6cm},centering}}

\newcommand{\nofancyheadfoot}{%
\fancyhead[L]{}%
\fancyhead[C]{}%
\fancyhead[R]{}%
\fancyfoot[L]{}%
\fancyfoot[C]{}%
\fancyfoot[R]{}%
\pagestyle{plain}%
}

\newcommand{\hyperrefpdf}{\usepackage[colorlinks=true,citecolor=red,linkcolor=blue,urlcolor=blue,pdftex]{hyperref}}
\hyperrefpdf

\newcommand{\skipline}[1][1]{\vspace*{#1\baselineskip}}
\newcommand{\itemsepex}[1][1]{\setlength{\itemsep}{#1ex}}

\usepackage[numbers,super,sort&compress]{natbib}
\afourlengths

\preauthor{\skipline[-1.75] \begin{center} \large}
\postauthor{\end{center}}
\predate{\skipline[-1.75] \begin{center} \large}
\postdate{\end{center} \skipline[-1.75]}

\hyperrefpdf

\nofancyheadfoot

\title{The Many Definitions of a Black Hole\thanks{Published in
    \emph{Nature Astronomy} 2019,
    \href{http://dx.doi.org/10.1038/s41550-018-0602-1}
    {doi:10.1038/s41550-018-0602-1} (free read-only SharedIt link:
    \url{https://rdcu.be/bfNpM}; preprint:
    \href{http://arxiv.org/abs/1808.01507} {arXiv:1808.01507
      [gr-qc]}).  An earlier version of this paper was titled ``What
    Is a Black Hole?''}}

\author{Erik Curiel$^{1,2,3}$}

\date{}

\begin{document}
\thispagestyle{empty}
\maketitle

\begin{affiliations}
    \item Munich Center for Mathematical Philosophy, 
  Ludwig-Maximilians-Universit\"at, Ludwigstra{\ss}e 31, 80539
  M\"unchen, Deutschland
      \item Black Hole
    Initiative, Harvard University, 20 Garden Street, Cambridge, MA
    02138, USA
      \item Smithsonian Astrophysical Observatory, Radio and
    Geoastronomy Division, 60 Garden Street, Cambridge, MA 02138, USA
\end{affiliations}




\begin{quote}
  \begin{center}
    \textbf{ABSTRACT}
  \end{center}  

  Although black holes are objects of central importance across many
  fields of physics, there is no agreed upon definition for them, a
  fact that does not seem to be widely recognized.  Physicists in
  different fields conceive of and reason about them in radically
  different, and often conflicting, ways.  All those ways, however,
  seem sound in the relevant contexts.  After examining and comparing
  many of the definitions used in practice, I consider the problems
  that the lack of a universally accepted definition leads to, and
  discuss whether one is in fact needed for progress in the physics of
  black holes.  I conclude that, within reasonable bounds, the
  profusion of different definitions is in fact a virtue, making the
  investigation of black holes possible and fruitful in all the many
  different kinds of problems about them that physicists consider,
  although one must take care in trying to translate results between
  fields.
\end{quote}

\skipline


\section*{The Question}
\label{sec:question}

What is a black hole?  That may seem an odd question.  Given the
centrality of black holes to theoretical work across many fields of
physics today, how can there be any uncertainty about it?  Black holes
(and their analogues) are objects of theoretical study in almost
everything from optics to solid-state to superfluids to ordinary
hydrodynamics and thermodynamics to high-energy particle physics to
astrophysics to cosmology to classical, semi-classical and quantum
gravity; and of course they are central subjects of observational work
in much of astronomy.  That fact perhaps provides part of the answer
about the uncertainty: there is not so much uncertainty about a
single, canonical answer, but rather there are too many good possible
answers to the question, not all consistent with each other.  That is
what makes the question of interest.  There is likely no other
physical system of fundamental importance about which so many
different answers are to be had for its definition, and so many
reasons to be both satisfied and dissatisfied with all of them.
Beatrice Bonga, a theoretical physicist, summed up the situation
admirably (personal communication): ``Your five word question is
surprisingly difficult to answer \ldots \space and I definitely won't
be able to do that in five words.''  (From hereon, when I quote
someone without giving a citation, it should be understood that the
source is personal communication.)

The question is not only interesting (and difficult) in its own right.
It is also important, both for practical reasons and for foundational
ones.  The fact that there are so many potentially good answers to it,
and seemingly little recognition across the fields that each relies on
its own peculiar definition (or small set of definitions), leads to
confusion in practice.  Indeed, I first began to think deeply about
the question when I noticed, time and again, disagreements between
physicists about what to my mind should have been basic points about
black holes all would agree on.  I subsequently traced the root of the
disagreements to the fact that the physicists, generally from
different fields (or even only different subfields within the same
field, such as different approaches to quantum field theory on curved
spacetime), were implicitly using their own definition of a black
hole, which did not square easily with that of the others in the
conversation.  Different communities in physics simply talk past each
other, with all the attendant difficulties when they try to make
fruitful contact with one another, whether it be for the purposes of
exploratory theoretical work, of concrete observational work, or of
foundational investigations.  (Ashtekar and
Krishnan\citep{ashtekar-krishnan-dyn-horiz-props} in a review of work
on isolated horizons give the only discussion I know in the literature
on this exact issue, that different fields of physics use different
definitions and conceptions of a black hole.)

The profusion of possible definitions raises problems that are
especially acute for foundational work.  The ground-breaking work of
Hawking\citep{hawking-bh-explode,hawking-part-create-bh} concluded
that, when quantum effects are taken into account, black holes should
emit thermalized radiation like an ordinary blackbody.  This appears
to point to a deep and hitherto unsuspected connection among our three
most fundamental, deeply entrenched theories, general relativity,
quantum field theory, and thermodynamics.  Indeed, black hole
thermodynamics and results concerning quantum fields in the presence
of strong gravitational fields more generally are without a doubt the
most widely accepted, most deeply trusted set of conclusions in
theoretical physics in which those theories work together in seemingly
fruitful harmony.  This is especially remarkable when one reflects on
the fact that we have absolutely no experimental or observational
evidence for any of it, nor hope of gaining empirical access any time
soon to the regimes where such effects may appreciably manifest
themselves.

All is not as rosy, however, as that picture may paint it.  Those
results come from taking two theories (general relativity and quantum
field theory), each of which is in manifest conceptual and physical
tension with the other in a variety of respects, and each of which is
more or less well understood and supported in its own physical regime
radically separated from that of the other, and attempting to combine
them in novel ways guided by little more than physical intuition
(which differs radically from physicist to physicist) and then to
extend that combination into regimes we have no hard empirical
knowledge of whatsoever.  It is far from clear, however, among many
other issues, what it may mean to attribute thermodynamical properties
to black holes.\citep{curiel-class-bhs-hot} \space The problem is made
even more acute when one recognizes that the attribution suffers of
necessity the same ambiguity as does the idea of a black hole itself.
Attempts to confront such fundamental problems as the Information-Loss
Paradox\citep{marolf-bh-info-loss-past-pres-fut,unruh-wald-info-loss}
are in the same boat.  Since almost all hands agree that black hole
thermodynamics provides our best guide for clues to a successful
theory of quantum gravity, it would be useful to know what exactly
those clues are.  Thus, it behooves us to try to get clear on what
black holes are.

I shall speak in this essay as though the task is to provide a
definition, in perhaps something like a logical sense, for black
holes.  In their daily practice, I suspect most physicists do not
think in those terms, having rather more or less roughly delineated
conceptions they rely on in their work, a picture of what they mean by
``black hole''.  Nonetheless, for ease of exposition, I will continue
to speak of definitions.


\section*{The History}
\label{sec:history}

In the 1960s, our understanding of general relativity as a theory
experienced a revolution at the hands of Penrose, Hawking, Geroch,
Israel, Carter and others, with the development of novel techniques in
differential topology and geometry to characterize the global
structure of relativistic spacetimes in ways not tied to the specifics
of particular solutions and independent of the assumption of high
degrees of symmetry.  This work in part originated with the attempt to
understand the formation of singularities and the development of the
causal structure of spacetime during the gravitational collapse of
massive bodies such as stars.  It culminated in the classic definition
of a black hole as an event horizon (the boundary of what is visible
from, and therefore what can in principle escape to, ``infinity''),
the celebrated singularity theorems of Penrose, Hawking, and Geroch,
the No-Hair theorems of Israel, Carter and others, Penrose's
postulation of the Cosmic Censorship Hypothesis, the demonstration
that trapped surfaces (close cousins to event horizons) will form
under generic conditions during gravitational collapse, and many other
results in classical general relativity that today ground and inform
every aspect of our understanding of relativistic spacetimes.  (For
those interested in the fascinating history of the attempts to
understand black-hole solutions to the Einstein field equation before
the 1960s, see Earman\citep{earman-bangs}, Earman and
Eisenstadt\citep{earman-eisenstaedt-einstein-sings}, and
Eisenstadt\citep{eisenstaedt-early-interp-schwarz}.)

Among the community of physicists steeped in classical general
relativity, exemplified by the groups associated with John Wheeler at
Princeton and Dennis Sciama at Cambridge, this was heady stuff.
According to active participants of those groups at the time, no one
in that community had the least doubt about what black holes were and
that they existed.  

It was otherwise with astrophysics and more traditional cosmology in
the 1960s.  There was controversy about whether or not to take
seriously the idea that black holes were relevant to real-world
physics.  For many, black holes were just too weird---according to the
relativists' definition, a black hole is a global object, requiring
that one know the entire structure of spacetime to characterize it
(more on this below), not a local object determinable by local
observations of phenomena of the sort that are the bread and butter of
astrophysics.  In his classic text on general relativity and
cosmology, Weinberg\citep{weinberg72}, for instance, strongly suggests
that black holes are not relevant to the understanding of compact
cosmological objects such as quasars, expresses deep skepticism that
real stars will collapse to within their Schwarzschild radius even
while citing Penrose on the formation of trapped surfaces, and
completely dismisses the idea that the interior of the event horizon
of the Schwarzschild black hole is relevant for understanding collapse
at all.

One crucial point that astrophysicists and cosmologists of the time
were not in a position to recognize, however, because of their
conception of black holes as a spatially localized, compact object
formed by collapse from which nothing can escape, is that black holes
are not associated only with traditional collapse phenomena.  As Bob
Geroch, a theoretical physicist known for his work in classical
general relativity, points out, if all the stars in the Milky Way
gradually aggregate towards the galactic center while keeping their
proportionate distances from each other, they will all fall within
their joint Schwarzschild radius long before they are forced to
collide.  There is, in other words, nothing necessarily unphysical or
mysterious about the interior of an event horizon formed from the
aggregation of matter.  Reasoning such as this based on their
definition of a black hole as a spacetime region encompassed by an
event horizon confirmed the relativists in their faith in the
existence of black holes, confirmation buttressed by the conviction,
based on Penrose's results about the formation of trapped surfaces
during generic collapse, that the extremity of self-gravitational
forces in traditional collapse would overwhelm any possible
hydromagnetic or quantum effects resisting it.

This paints the picture with an extremely broad and crude brush, and
there were many astrophysicists and cosmologist who did not conform to
it.  As early as 1964 Edwin Salpeter and Yakov Zel'dovi{\^c} had
independently argued that supermassive black holes accreting gas at
the centers of galaxies may be responsible for the enormous amounts of
energy emitted by quasars, along with their large observed variability
in luminosity.  In the early 1970s, Donald Lynden-Bell proposed that
there is a supermassive black hole at the center of the Milky Way.
Zel'dovi{\^c} in Moscow and groups led by Lynden-Bell and by Martin
Rees in Cambridge (UK) at the same time independently worked out
detailed theoretical models for accretion around black holes for
quasars and x-ray binaries.

Based on observational work, astrophysicists knew that some massive,
compact object had to be at the center of a quasar, but there was
still reticence to fully embrace the idea that it was a black hole.
Accretion onto a black hole was at that point the widely accepted
model, to be sure, but the seemingly exotic nature of black holes left
many astrophysicists with unease; there was, however, no other
plausible candidate known.  With upper possible mass limits on neutron
stars worked out in the 1970s, and more and more observational
evidence coming in through the 1980s that the objects at the center of
quasars had to be more massive than that, and compressed into an
extremely small volume, more and more doubters were won over as
theoretical models of no other kind of system could so well account
for it all.  (It is amusing to note, however, that even well into the
1980s Bob Wald, a theoretical physicist at the University of Chicago,
had to warn astrophysicists and cosmologists visiting there against
describing black holes as ``exotic'' in their talks, as that would
have led to the interruption of their talk for chastisement by
Chandrasekhar.)  Cygnus X-1 and other X-ray binaries also provided
observational evidence for black holes in the early 1970s.  It is
perhaps fair to say that the community achieved something like
unanimous agreement on the existence and relevance of black holes only
in the early 2000s, with the unequivocal demonstration that SgrA$^*$,
the center of the Milky Way, holds a supermassive black hole, based on
a decade of infrared observations by Reinhard Genzel, Andreas Eckart,
and Andrea
Ghez\citep{genzel-et-dark-mass-ctr-milky,ghez-et-acc-stars-orbit-bh}.


\section*{Possible Answers}
\label{sec:poss-answers}

In \emph{Confessions}, Saint Augustine famously remarked, ``Quid est
ergo tempus?  Si nemo ex me qu{\ae}rat, scio; si qu{\ae}renti
explicare velim, nescio.'' (``What then is time?  If no one asks me, I
know what it is.  If I wish to explain it to someone who asks, I do
not know.\@'' Lib.~\textsc{ix}, cap.~14.)  As for time, so for black
holes.  Most physicists, I believe, know what a black hole is, right
up until the moment you blindside them with the request for a
definition.  In preparation for writing this essay, I did exactly
that.  I posed the question, with no warning or context, to physicists
both young and old, just starting out and already eminent,
theoretician and experimentalist, across a wide variety of fields.
The results were startling and eye-opening, not only for the variety
of answers I got but even more so for the puzzlement and deep
thoughtfulness the question occasioned.

I will discuss the possible definitions in detail shortly.  Before
diving in, however, it will be useful to sketch the terrain in rough
outline.  In table~\ref{tab:core-concepts}, I lay out the core
concepts that workers in different fields tend to use when thinking
about black holes.  The table, however, is \emph{only} a rough guide.
As we can see from the quotes from physicists in different fields
given in separate boxes at the end of the essay, and from the more
detailed discussion below, not all physicists in a given field conform
to the standard.
\begin{table}
  \centering
  \begin{tabular}[c]{|p{.4\textwidth}|p{.6\textwidth}|}
    \hline
    \multicolumn{1}{|c|}{\textbf{Field}} & \multicolumn{1}{|c|}{\textbf{Core Concepts}} \\
    \hline
    \skipline[.5] astrophysics &  \skipline[-1]
    \begin{itemize}
      \itemsepex[-1]
        \item compact object
        \item region of no escape
        \item engine for enormous power output \skipline[-.5]
    \end{itemize} \\
    \hline 
    \skipline[1] classical relativity & \skipline[-1]
    \begin{itemize}
      \itemsepex[-1]
        \item causal boundary of the past of future null infinity
      (event horizon)
        \item apparent horizon
        \item quasi-local horizon \skipline[-.5]
    \end{itemize} \\
    \hline 
    \skipline[.1] mathematical relativity & \skipline[-1]
    \begin{itemize}
      \itemsepex[-1]
        \item apparent horizon
        \item singularity \skipline[-.5]
    \end{itemize} \\
    \hline 
    \skipline[.5] semi-classical gravity & \skipline[-1]
    \begin{itemize}
      \itemsepex[-1]
        \item same as classical relativity
        \item thermodynamical system of maximal entropy
      \skipline[-.5]
    \end{itemize} \\
    \hline 
    \skipline[1] quantum gravity & \skipline[-1]
    \begin{itemize}
      \itemsepex[-1]
        \item particular excitation of quantum field
        \item ensemble or mixed state of maximal entropy
        \item no good definition to be had \skipline[-.5]
    \end{itemize} \\
    \hline
    \skipline[.15] analogue gravity & \skipline[-.75]
    \begin{itemize}
      \itemsepex[-1]
        \item region of no escape for finite time, or for low energy
      modes \skipline[-.5]
    \end{itemize} \\
    \hline
  \end{tabular}
  \caption{\textbf{The core concepts common to different fields
      for characterizing black holes.}}
  \label{tab:core-concepts}
\end{table}
  
Most likely because of my training and the focus of most of my own
work in classical general relativity and semi-classical gravity, I
naively expected almost everyone I asked at least to mention ``the
boundary of the causal past of future null infinity'', the classic
definition of the event horizon dating back to the ground-breaking
work of the mid-to-late 1960s, as laid down in the canonical texts on
general relativity by Hawking and
Ellis\citep{hawking-ellis-lrg-scl-struc-st} and by
Wald\citep{wald-gr}.  In the event, many did not, and most of those
who mentioned it did so at least in part to draw attention to its
problems.  The definition tries to take the intuition that a black
hole is a ``region of no escape'' and make it precise.  In order for
the idea of a region of no escape to be cogent, there must be another
region stuff possibly could escape to, so long as it never enters the
trapping region.  The definition thus states in effect that a
spacetime has a black hole if one can divide the spacetime into two
mutually exclusive, exhaustive regions of the following kinds.  The
first, the exterior of the black hole, is characterized by the fact
that it is causally connected to a region one can think of as being
``infinitely far away'' from the interior of the spacetime; anything
in that exterior region can, in principle, escape to infinity.  The
second region, the interior of the black hole, is characterized by the
fact that once anything enters it, it must remain there and cannot,
not even in principle, escape to infinity, nor even causally interact
in any way with anything in the other region.  The boundary between
these two regions is the event horizon.


This definition is global in a strong and straightforward sense: the
idea that nothing can escape the interior of a black hole once it
enters makes implicit reference to all future time---the thing can
never escape \emph{no matter how long it tries}.  Thus, in order to
know the location of the event horizon in spacetime, one must know the
entire structure of the spacetime, from start to finish, so to speak,
and all the way out to infinity.  As a consequence, no local
measurements one can make can ever determine the location of an event
horizon.
That feature is already objectionable to many physicists on
philosophical grounds: one cannot operationalize an event horizon in
any standard sense of the term.  Another disturbing property of the
event horizon, arising from its global nature, is that it is
prescient.  Where I locate the horizon today depends on what I throw
in it tomorrow---which future-directed possible paths of particles and
light rays can escape to infinity starting today depends on where the
horizon will be tomorrow, and so that information must already be
accounted for today.  Physicists find this feature even more
troubling.

\skipline

\begin{center}
  \begin{tabular}[c]{|p{.9\textwidth}|}
    \hline
    \begin{quote}
      The existence of [a classical event horizon] just doesn't seem
      to be a verifiable hypothesis.
      \skipline[-.5]
      \begin{flushright}
        -- Sean Gryb, theoretical physicist \\
        (shape dynamics, quantum cosmology)
      \end{flushright}
    \end{quote}
    \\
    \hline
  \end{tabular}
\end{center}

\skipline

For reasons such as those, some physicists define a black hole as a
kind of horizon whose characteristic properties may be relative to a
particular set of observers and their investigative purposes, similar
to how ``equilibrium'' in thermodynamics must be defined for a system
with respect to some characteristic time-scale picked out by the
physics of the problem at hand.  Other physicists propose
generalizing the classic definition in other ways that make explicit
reference to observers, so-called causal
horizons.\citep{jacobson-parentani-horiz-ent} \space This allows one
to bring the concept of a black hole as a horizon into immediate
contact with other more general kinds of horizons that appear in
general relativity, in order to formulate and prove propositions of
great scope about, say, their thermodynamical properties.  It is
interesting to note that several of these other conceptions of a
horizon do not depend on a notion of infinity in the sense of a place
one can unambiguously escape to (null or spatial infinity), but they
do still make implicit reference to a future temporal infinity.

Such causal horizons are still global in nature, however, so, in
attempting to assuage the general dissatisfaction with the global
nature of
the classic definition, one possible strategy is to attempt to isolate
some characteristic feature of a global black hole that can be
determined locally.  One popular such feature is a so-called
\emph{apparent horizon}, a structure that generically appears along
with a classical event horizon, but whose existence and location can
seemingly be determined locally, and which can also be defined in
spacetimes in which an event horizon cannot, \emph{e}.\emph{g}., those
that are bounded in space so there is no good notion of ``escape to
infinity''.  An apparent horizon is a two-dimensional surface (which
we may for our purposes think of as a sphere) such that, loosely
speaking, all light rays emanating outward from nearby points on its
surface start out parallel to each other.  This captures the idea that
``nothing, not even light, can escape'' in a local fashion---outgoing
light wants to remain tangent to the surface.  Note, however, that
there is no guarantee that something entering the region bounded by a
suitable characterization of the future evolution of such a surface
may not later be able to exit from it.

Many characteristic properties of classical event horizons follow
already from the idea of an apparent horizon, and it is easily
generalized to alternative theories of gravity (\emph{i}.\emph{e}.,
non-quantum gravitational theories that differ from general
relativity).  Nonetheless, apparent horizons (and other such ``local''
notions of a horizon, which I discuss briefly below) are not quite so
local as commonly held opinion assumes: to determine that a surface is
an apparent horizon, one still needs to determine that neighboring
outgoing light rays propagate parallel to each other \emph{all at once
  on the entire surface}.  No observer could ever determine this in
practice, though perhaps a large team of perfectly synchronized
observers could do it in principle.  An even more serious problem,
however, is that apparent horizons are slice-dependent,
\emph{i}.\emph{e}., whether one takes an apparent horizon to be
present or not depends on how one foliates spacetime by spacelike
hypersurfaces---on how one locally splits spacetime up into spatial
and temporal parts.  Many physicists are uncomfortable with grounding
reasoning of a fundamental nature on objects or structures that are
not invariantly defined with respect to the full 4-dimensional
spacetime geometry.

Mathematicians in general are also leery of the global nature of the
classic definition.  In recent decades, mathematical relativity has
largely focused on studying the initial-value problem of general
relativity, attempting to
characterize solutions to the Einstein field equation viewed as a
result of dynamical evolution starting from initial data on
3-dimensional spacelike hypersurfaces.  This initial data determines
spacetime structure locally in the domain of evolution.  Because the
presence of apparent horizons can be determined locally in a
mathematically relevant sense, they often use this as the marker that
a black hole is present.  Under a few seemingly benign assumptions,
moreover, the presence of an apparent horizon leads by the classic
Penrose singularity theorem\citep{penrose-grav-coll-st-sings} to the
existence of a singularity one expects to find inside a black hole.
Since the presence of a singularity can also be determined locally, it
is often included in the definition of a black hole for
mathematicians.

The mathematicians' conception does not, however, meet all their
own desiderata.  First, the initial data is not truly local---one must
in general specify conditions on it asymptotically, at ``spatial
infinity'', and it is difficult at best to see why needing to know the
structure of spacetime at ``all of space at a given moment of time''
is epistemically superior to needing to know the future structure of
spacetime.  Even worse, it does not suffice for an unambiguous
definition of a black hole.  We have little understanding of the
evolution of generic initial data for the Einstein field equation.  We
know of no way in general to determine whether a set of locally
stipulated initial conditions will eventuate in anything like a
classical horizon or singularity, except by explicitly solving the
equations, and that is almost never feasible in practice, outside
special cases of unrealistically high degrees of symmetry.

\skipline

\begin{center}
  \begin{tabular}[c]{|p{.9\textwidth}|}
    \hline
    \begin{quote}
      [The classic conception of a horizon] is probably a very useless
      definition, because it assumes we can compute the future of real
      black holes, and we cannot.  \skipline[-.5]
      \begin{flushright}
        -- Carlo Rovelli, theoretical physicist \\
        (classical general relativity, loop quantum gravity,
        cosmology, foundations of quantum mechanics)
      \end{flushright}
    \end{quote}
    \\
    \hline
  \end{tabular}
\end{center}

\skipline

Besides the apparent horizon, there are other quasi-local
characterizations of black holes that do not have objectionable global
features, such as dynamical trapping
horizons\citep{hayward-genl-laws-bhdyns} and isolated
horizons\citep{ashtekar-et-isol-horiz}.  Several physicists and
astrophysicists in their replies to me mention these, mainly to
discuss their virtues, but they are difficult to describe without
resorting to technical machinery.  One may usefully think of them as
closed surfaces that have many of the properties of apparent horizons,
without necessarily being associated with a classical event horizon.
They have problems of their own, though, a severe one being that they
are
slice-dependent in the same way as apparent horizons.  Also, perhaps
even worse, they have a form of ``clairvoyance'': they are aware of
and respond to changes in the geometry in spacetime regions that they
cannot be in causal contact
with\citep{bengtsson-senovilla-trpd-surfs-sphl-sym}.  Indeed, they can
encompass regions whose entire causal past is flat.  This should be at
least as troubling as the ``prescience'' of global event horizons.

The global and prescient nature of the classical event horizon never
bothered me.  I see the classic definition as an elegant and powerful
idealization, nothing else, allowing us to approximate the spacetime
structure around a system that is for all intents and purposes
isolated from the rest of the universe in the sense that the
gravitational (and other) effects of all other systems are
negligible---spacetime in our neighborhood \emph{is} approximately
flat compared to regions around objects we attempt to study and think
of as black holes, and we are very, very far away from them.  It is
also an idealization that allows us to prove theorems of great depth
and scope, giving us unparalleled insight into the conceptual
structure of general relativity as a physical theory (in so far as one
trusts results based on the idealization to carry over to the real
world).  This of course still leaves us with the task of
characterizing what it means for a region of spacetime to ``act
approximately like a black hole'' in a way that renders the
idealization suitable for our purposes.  Given the number of features
one may want to take as characteristic and try to hold on to, and the
fact that one will not be able to hold on to all of them (as discussed
below), this still leaves a great deal of freedom in fleshing
out the idea of ``acting approximately like a
black hole'' as a fruitful conception, and that presumably will again
depend on the details of the investigations at hand and the purposes
of the physicists engaged in them.

Astrophysicists, in their applied work, tend to be sanguine about the
global nature of the classic definition.  They are happy to avail
themselves of the deep results about horizons that the classic
definition allows us to prove when, \emph{e}.\emph{g}., they try to
determine
what observable properties a region of spacetime may have that would
allow us to conclude that what we are observing is a black hole in
their sense.  They still use in their ordinary practice, nonetheless,
a definition that is tractable for their purposes: a system of at
least a minimum mass, spatially small enough that relativistic effects
cannot be ignored.  Neutron stars cannot have mass greater than about
3 solar masses, and a star with greater mass will not be relativistic
in the relevant sense.  It more or less follows from this, as other
astrophysicists stress as a characteristic property when defining a
black hole, that it be a region of no escape in a sense relevant to
their work.

\skipline

\begin{center}
  \begin{tabular}[c]{|p{.9\textwidth}|}
    \hline
    \begin{quote}
      A black hole is a compact body of mass greater than 4 Solar
      masses---the physicists have shown us there is nothing else it
      can be.  \skipline[-.5]
      \begin{flushright}
        -- Ramesh Narayan, astrophysicist \\
        (active galactic nuclei, accretion disc flow)
      \end{flushright}
    \end{quote}
    \\
    \hline
  \end{tabular}
\end{center}

\skipline

None of this, however, distinguishes a black hole from a naked
singularity (\emph{i}.\emph{e}., a singularity not hidden behind an
event horizon, ruled out by Penrose's Cosmic Censorship
Conjecture\citep{penrose-grav-coll-role-gr}).  Astrophysicists tend to
respond to this problem in two ways.  First, they try to exclude the
possibility of naked singularities on other theoretical grounds;
second, much work is currently being done to try to work out
properties of naked singularities that would distinguish them
observationally from black
holes\citep{narayan-mcclintock-obs-evid-bhs}.  There are many other
fascinating methodological and epistemological problems with trying to
ascertain that what we observe astronomically conforms to these sorts
of
definitions,\citep{collmar-et-panel-proof-exist-bhs,eckart-et-superm-bh-good-case}
but it would take us too far afield to go into them here.

It is worth remarking that it is not only astrophysicists who share
this conception.  Many theoretical physicists working in programs from
high-energy particle physics to loop quantum gravity also champion
definitions that latch on to one facet or another of the standard
astrophysics definition.
Gerard 't Hooft, for instance, in his remarks quoted at the end of the
essay, emphasizes his conception of a black hole as a vacuum solution
resulting from total collapse, adding a subtle twist to the
astrophysicist's concrete picture in which ordinary matter may be
present (\emph{e}.\emph{g}., in an accretion disc), a twist perhaps
congenial to a particle physicist's aims of investigating the
transformations of the vacuum state of a quantum field in the vicinity
of a
horizon.  Others take over the astrophysicist's picture wholesale,
emphasizing that previous purely theoretical conceptions are no longer
adequate for contemporary work that would make contact with real
observations, as Carlo Rovelli makes clear in his remarks quoted at
the end.  Nonetheless, as well as the astrophysicist's picture may
work in practice, it also faces serious conceptual problems.  Black
holes simply are not anything like other kinds of astrophysical
systems that we study---they are not bits of stuff with
well defined spatiotemporal positions that interact with ordinary
systems in a variety of ways other than gravity.

In the semi-classical framework, one treats the spacetime geometry as
classical, with quantum fields propagating against it as their
background.  In that picture, some of the concerns just discussed
appear to be mitigated.  Black holes seem to acquire some of the most
fundamental properties of ordinary physical systems: they exhibit
thermodynamical behavior.  The presence of Hawking radiation, a
consequence of the semi-classical approach, allows us to define a
physical temperature for a black hole\citep{wald-grav-thermo-qt}.
Semi-classical proofs of the Generalized Second Law, moreover, justify
the attribution of entropy to a black hole proportional to its
area\citep{wall-10-proofs-gsl}.  In the standard semi-classical
picture,
moreover, most researchers hold that the classical characterizations
of black holes are unproblematic (or, at least, no more problematic
than in the strictly classical context).  The geometry is classical,
they reason, so we can avail ourselves of all the tools we use to
characterize black holes in the classical regime.  Nonetheless, in so
far as we do accept the semi-classical picture of black holes
evaporating as they emit Hawking radiation, we must give up entirely
on the idea of black holes as eternal, global objects, and use that
idealization with care.  The very presence of Hawking radiation
itself, furthermore, independently of the role it may play in black
hole evaporation, means that we may also need to give up on the
classical idea of black holes as perfect absorbers, and all the many
important consequences that property entails.

\skipline

\begin{center}
  \begin{tabular}[c]{|p{\textwidth}|}
    \hline
    \begin{quote}
      If we do accept the picture [of semi-classical gravity], then
      black holes become for the first time now, in this context, true
      physical systems---they have thermodynamical properties.
      \skipline[-.5]
      \begin{flushright}
        -- Daniele Oriti, theoretical physicist \\
        (semi-classical gravity, group-field theory quantum gravity)
      \end{flushright}
    \end{quote}
    \\
    \hline
  \end{tabular}
\end{center}

\skipline

That, however, is a claim it is delicate to make precise, exactly
because of the subtle interplay between the quantum effects of matter
and the classical geometry.  It is difficult to say with precision and
clarity whether or not Hawking radiation shows that the interior of a
black hole cannot be wholly isolated causally from its exterior.  That
ambiguity, however, calls into question the very distinction between
the interior and the exterior of a black hole that the idea of an
event horizon is supposed to explicate.  I believe the idea of a black
hole in the semi-classical context is not so clear cut as almost all
physicists working in the field seem to think.  Indeed, that black
holes seem to have a non-trivial thermodynamics pushes us towards the
view that there is an underlying dynamics of micro-degrees of freedom
that is not and seemingly cannot be captured in the semi-classical
picture, perhaps undermining the very framework that suggested it in
the first place.  In the same vein, it is well to keep in mind that
none of the results in the semi-classical domain about black hole
thermodynamics come from fundamental theory, but rather from a
patchwork of different methods based on different intuitions and
principles.  As I mentioned already in the introduction, the
semi-classical picture comes from trying to combine in completely
novel ways two theories that are in manifest tension with each other,
absent the guidance and constraint of experimental or observational
knowledge.  I think it behooves us to show far more caution in
accepting the results of semi-classical black hole thermodynamics than
is common in the field.

In other approaches with a semi-classical flavor, such as the
conjectured duality between gravitational physics in anti-de{} Sitter
spacetime and conformal field theories on its boundary
(AdS-CFT)\citep{maldacena-sup-conf-fld-sup-grav}, and many projects
based on holography more
generally,\citep{thooft-dim-reduct-qgrav,thooft-holo-princ,bousso-holo-gen-backgrnd}
it is difficult to define black holes at all in any direct way.  In
such approaches, one posits that the classical gravitational physics
in an interior region of a spacetime is entirely captured by the
physics of a quantum field on the boundary of the region (the timelike
boundary at infinity in anti-de{} Sitter spacetime,
\emph{e}.\emph{g}.).  It is not easy to read off from the boundary
physics whether anything resembling a black hole in any of its many
guises (a horizon of a particular sort, for instance) resides in the
interior.

There are attempts to do so, however, by isolating characteristic
features of the configuration and evolution of the quantum fields on
the boundary associated with black hole spacetimes in the interior.
The holographic principle would then suggest that one identify those
field configurations having maximal entropy as black holes.  In a
similar vein, some physicists working in holography and string theory,
such as Juan Maldacena (personal communication), suggest that one
characteristic feature of black holes is that their dynamical
evolution is maximally chaotic, part and parcel of their purported
entropy-maximization properties\citep{maldacena-et-bound-chaos}.
Others, such as 't Hooft (personal communication), reject that idea,
contending that the main gravitational effect that governs how black
holes behave is completely linear, and so they cannot serve as
information scramblers in the sense championed by many others in the
holography community.  One physicist's characteristic property is
another's mistaken claim.

Even if one does accept any of the glosses available in holography,
one must face the fact that it is difficult to extract from the
physics of the boundary field anything about the physics of the
interior of a classical event horizon, a well known problem in these
approaches.  Any definition that cannot handle the interior of a black
hole, however, must have a demerit marked against it.  No known
quantum effect, nor any other known or imagined physical process, can
cause spacetime simply to stop evolving and vanish, as it were, once
matter crosses its Schwarzschild radius.  Perhaps nothing inside a
horizon can communicate with the outside, but that does not mean it is
not part of the world.  As such, the mettle of physics demands that we
try to understand it.

In quantum gravity in general, most agree that the problems of
defining a black hole in a satisfactory manner become even more
severe.
There is, for instance, in most programs of quantum gravity, nothing
that corresponds to an entire classical history
on which to base something like the traditional definition.  Even
trying to restrict oneself to quasi-local structures such as the
apparent horizon has manifest problems: in the quantum context, in
order to specify the geometry of such a surface, one in effect has to
stipulate simultaneously values for the analogues of both the position
and momentum of the relevant micro-structure, a task that quantum
mechanics strongly suggests cannot be coherently performed.

\skipline

\begin{center}
  \begin{tabular}[c]{|p{.9\textwidth}|}
    \hline
    \begin{quote}
      Ideally the definition used in Quantum Gravity reduces to the
      one in classical General Relativity in the limit $\hbar$ goes to
      zero\ldots.\space\space But since no one agrees on what a good
      theory of quantum gravity is (not even which principles it
      should satisfy), I don't think anyone agrees on what a black
      hole is in quantum gravity.  \skipline[-.5]
      \begin{flushright}
        -- Beatrice Bonga, theoretical physicist \\
        (gravitational radiation, quantum gravity phenomenology)
      \end{flushright}
    \end{quote}
    \\
    \hline
  \end{tabular}
\end{center}

\skipline

One strategy for characterizing a black hole common to many approaches
to quantum gravity is to ask, what particular kind of ensemble or
assembly of building blocks constructed from the fundamental degrees
of freedom ``looks like'' a black hole, when one attempts to impose on
them in some principled way a spatiotemporal or geometrical
``interpretation''?  The idea is to try to put together ``parts'' of
the classical picture of a black hole one by one---find properties of
an underlying quantum ensemble that make the resulting ``geometry''
look spherically symmetric, say, and make it amenable to having a
canonical area attributed to it, and so on, building up to the
semi-classical picture\citep{oriti-et-bhs-qg-condens}.  It is
difficult to test the conjecture that this will correspond to a
classical black hole, however, in any known program of quantum
gravity, because it is difficult to reconstruct the causal structure
of the ``resulting'' classical geometry.  A related strategy that
suggests itself, inspired by the holographic principle, is to put
together a quantum ensemble that in some sense is sharply peaked
around a spherically symmetric geometry at the semi-classical level, a
geometry moreover that respects the quasi-local conditions imposed by
the classical picture of what a horizon should be.  One then attempts
to compute the entropy, maximizes it, and finally declares that the
resulting ensemble is the \emph{definition} of a black hole.  The
conjecture that this corresponds to a classical black hole is, again,
difficult to verify theoretically, and of course impossible at the
present time to test by experiment, and will be so for the foreseeable
future.



Finally, although stricitly speaking not work in gravitational
physics, it is of interest to look briefly at so-called analogue
models of
gravity\citep{unruh-dumb-holes-anlg-bhs,jacobson-bhs-hawk-rad-st-anlgs}.
The explosion of work in that field centers on generalizations of the
idea of a black hole, in the guise of a horizon of an appropriate sort
across a broad range of non-gravitational types of physical systems.
The kinds of horizon at issue here will of necessity be
generalizations in some sense of the kinds one finds in relativity,
since one does not have available here the full toolbox of classical
spacetime geometry to work with.  The fundamental problem is that the
horizons one deals with in analogue systems are never true one-way
barriers.
This raises fascinating problems about how much or even whether at all
one should trust the results of experimental and theoretical work in
that field to translate into confirmatory support for the
semi-classical gravity systems they are analogue models
of\citep{unruh-schutzhold-univ-hawking-eff,dardashti-et-conf-anlg-sim}.
Sadly, space does not permit discussing those problems here.

\section*{Why It Matters}
\label{sec:why-matters}

I believe there is a widespread hope across the many fields of physics
in which black holes are studied that, though the conceptions,
pictures, and definitions used differ in manifestly deep and broad
ways, nonetheless they are all at bottom trying to get at the same
thing.  It is difficult otherwise to see
how work in one area is to make fruitful contact with work in all the
other areas.  It is, however, at this point only a hope.  Much work
must be done to make clear exactly how all those different
definitions, characterizations, and conceptions relate to each other,
so we can have confidence when we attempt to apply results from one
field to problems in another.  That is why the question matters.

Consider Hawking radiation.  It is a problem oddly unremarked in the
literature that, in the semi-classical picture, Hawking radiation is
not blackbody radiation in the normal sense.  Blackbody radiation,
such as the electromagnetic radiation emitted by a glowing lump of hot
iron, is generated by the dynamics of the micro-degrees of freedom of
the system itself---in the case of iron, the wiggling and jiggling of
the iron's own atoms and free electrons that makes them radiate.  That
is not the mechanism by which Hawking radiation is produced.  In the
semi-classical picture, Hawking radiation is not generated by the
dynamics of any micro-degrees of freedom of the black hole itself, but
rather by the behavior of an external quantum field in the vicinity of
the horizon.
The hope, presumably, is that a satisfactory theory of quantum gravity
will be able to bring these two \emph{prima facie} disparate
phenomena---the horizon on the one hand, and the dynamics of the
external quantum field on the other---into explicit and harmonious
relation with each other so as to demonstrate that the temperature of
the thermalized quantum radiation is a sound proxy for the temperature
of the black hole itself as determined by the dynamics of its very own
micro-degrees of freedom.  Since Hawking radiation is universally
viewed as the strongest evidence in favor of attributing a temperature
to black holes, and so attributing thermodynamical properties more
generally to them, the lack of such an explicit connection ought to be
troubling.  It ought to become even more troubling when one considers
the difficulties of defining black holes in all the different relevant
contexts, and relating those different definitions in rigorous, clear,
precise ways.  How can the physicists across different fields hope to
agree on an answer when they do not even agree on the question?

\skipline

\begin{center}
  \begin{tabular}[c]{|p{\textwidth}|}
    \hline
    \begin{quote}
      You [Curiel] suggest that it should be troubling that black hole
      temperature seems very different from the temperature of
      ordinary matter.  I find this very intriguing and exciting, not
      troubling.  \skipline[-.5]
      \begin{flushright}
        -- Bob Wald, theoretical physicist \\
        (classical general relativity, quantum field theory on curved
        spacetime)
      \end{flushright}
    \end{quote}
    \\
    \hline
  \end{tabular}
\end{center}

\skipline

I suspect there will never be a single definition of ``black hole''
that will serve all investigative purposes in all fields of physics.
I think the best that can be done, rather, is, during the course of an
investigation, to fix a list of important, characteristic properties
of and phenomena associated with black holes required for one's
purposes in the context of interest, and then to determine which of
the known definitions imply the members of that list.  If no known
definition implies one's list, one either should try to construct a
new definition that does (and is satisfactory in other ways), or else
one should conclude that there is an internal inconsistency in one's
list, which may already be of great interest to learn.  Here are
potentially characteristic properties and phenomena some subset of
which one may require or want: \skipline[-1.25]
\begin{itemize}
  \itemsepex[-1]
    \item possesses a horizon that satisfies the four laws of
  black hole mechanics;
    \item possesses a locally determinable horizon; 
    \item possesses a horizon that is, in a suitable sense, vacuum;
    \item is vacuum with a suitable set of symmetries;
    \item defines a region of no escape, in some suitable sense, for
  some minimum period of time;
    \item defines a region of no escape for all time; 
    \item is embedded in an asymptotically flat spacetime; 
    \item is embedded in a topologically simple spacetime;
    \item encompasses a singularity; 
    \item satisfies the No-Hair Theorem; 
    \item is the result of evolution from initial data satisfying an
  appropriate Hadamard condition (stability of evolution);
    \item allows one to predict that final, stable states upon
  settling down to equilibrium after a perturbation correspond, in
  some relevant sense, to the classical stationary black hole
  solutions (Schwarzschild, Kerr, Reissner-Nordstr\"om, Kerr-Newman);
    \item agrees with the classical stationary black hole solutions
  when evaluated in those spacetimes;
    \item allows one to derive the existence of Hawking radiation from
  some set of independent principles of interest;
    \item allows one to calculate in an appropriate limit, from some
  set of independent principles of interest, an entropy that accords
  with the Bekenstein entropy (\emph{i}.\emph{e}., is proportional to
  the area of a relevant horizon, with corrections of the order of
  $\hbar$);
    \item possesses an entropy that is, in some relevant sense,
  maximal;
    \item has a lower-bound on possible mass;
    \item is relativistically compact.
\end{itemize}
\skipline[-1.25] This list is not meant to be exhaustive.  There are
many other such properties and phenomena one might need for one's
purposes.  It is already clear from this partial list, however, that
no single definition can accommodate all of them.  It is also clear
from the discussion that, even within the same communities, different
workers will choose different subsets of these properties for
different purposes in their thinking about black holes.

One may conclude that there simply is no common conceptual core to the
pre-theoretical idea of a black hole, that the hopeful conjecture that
physicists in different fields all refer to the same entity with their
different definitions has been thrown down on the floor and danced
upon.  I would not want to draw that conclusion, though neither do I
want to wholly endorse the strong claim that there is a single entity
behind all those multifarious conceptions.  I would rather say that
there is a rough, nebulous concept of a black hole shared across
physics, that one can explicate that idea by articulating a more or
less precise definition that captures in a clear way many important
features of the nebulous idea, and that this can be done in many
different ways, each appropriate for different theoretical,
observational, and foundational contexts.  I do not see this as a
problem, but rather as a virtue.  It is the very richness and
fruitfulness of the idea of a black hole that leads to this
multiplicity of different definitions, each of use in its own domain.
I doubt the idea would be so fruitful across so many fields if they
all were forced to use a single, canonical definition.

\newpage

\section*{Box 1: Astrophysical Views on Black Holes}

\begin{tabular}[c]{|p{\textwidth}|}
  \hline
  \begin{quote}
    A black hole is the ultimate prison: once you check in, you can
    never get out.
    \skipline[-.5]
    \begin{flushright}
      -- Avi Loeb, astrophysicist \\
      (cosmology, black hole evolution, first stars)
    \end{flushright}
  \end{quote}
  \\
  \hline
  \begin{quote}
    For all intents and purposes we \emph{are} at future null infinity
    with respect to SgrA$^*$.
    \skipline[-.5]
    \begin{flushright}
      -- Ramesh Narayan, astrophysicist \\
      (active galactic nuclei, accretion disc flow)
    \end{flushright}
  \end{quote}
  \\
  \hline
  \begin{quote}
    [I]n practice we don't really care whether an object is
    `precisely' a black hole.  It is enough to know that it acts
    approximately like a black hole for some finite amount of
    time\ldots.  [This is] something that we can observe and test.
    \skipline[-.5]
    \begin{flushright}
      -- Don Marolf, theoretical physicist \\
      (semi-classical gravity, string theory, holography)
    \end{flushright}
  \end{quote}
  \\
  \hline
  \begin{quote}
    [A black hole is] a region which cannot communicate with the
    outside world for a long time (where `long time' depends on what I
    am interested in).  \skipline[-.5]
    \begin{flushright}
      -- Bill Unruh, theoretical physicist \\
      (classical general relativity, quantum field theory on curved
      spacetime, analogue gravity)
    \end{flushright}
  \end{quote}
  \\
  \hline
\end{tabular}

\noindent\begin{tabular}[c]{|p{\textwidth}|}
  \hline
  \begin{quote}
    Today `black hole' means those objects we see in the sky, like for
    example Sagittarius A$^*$.
    \skipline[-.5]
    \begin{flushright}
      -- Carlo Rovelli, theoretical physicist \\
      (classical general relativity, loop quantum gravity, cosmology,
      foundations of quantum mechanics)
    \end{flushright}
  \end{quote}
  \\
  \hline
\end{tabular}

\skipline

\section*{Box 2: Classical Relativity and Semi-Classical Gravity Views
  on Black Holes}

\begin{tabular}[c]{|p{\textwidth}|}
  \hline
  \begin{quote}
    I'd \ldots \space define a causal horizon as the boundary of the
    past of an infinite timelike curve [\emph{i}.\emph{e}., the past
    of the worldline of a potential observer], and the black hole [for
    that observer] as the region outside the past.
    \skipline[-.5]
    \begin{flushright}
      -- Ted Jacobson, theoretical physicist \\
      (classical general relativity, semi-classical gravity, entropic
      gravity)
    \end{flushright}
  \end{quote}
  \\
  \hline
  \begin{quote}
    We [mathematicians] view a black hole to be a natural singularity
    for the Einstein equation, a singularity shielded by a membrane[,
    \emph{i}.\emph{e}., a horizon].
    \skipline[-.5]
    \begin{flushright}
      -- Shing-Tung Yau, mathematician, mathematical physicist \\
      (classical relativity, Yang-Mills theory, string theory)
    \end{flushright}
  \end{quote}
  \\
  \hline
  \begin{quote}
    A black hole is the solution of Einstein's field equations for
    gravity without matter, which you get after all matter that made
    up a heavy object such as one or more stars, implodes due to its
    own weight.
    \skipline[-.5]
    \begin{flushright}
      -- Gerard 't Hooft, theoretical physicist \\
      (Standard Model, renormalizability, holography)
    \end{flushright}
  \end{quote}
  \\
  \hline
\end{tabular}

\noindent\begin{tabular}[c]{|p{\textwidth}|}
  \hline
  \begin{quote}
    I have no idea why there should be any controversy of any kind
    about the definition of a black hole.  There is a precise, clear
    definition in the context of asymptotically flat spacetimes, [an
    event horizon]\ldots.\space\space I don't see this as any
    different than what occurs everywhere else in physics, where one
    can give precise definitions for idealized cases but these are not
    achievable/measurable in the real world.  \skipline[-.5]
    \begin{flushright}
      -- Bob Wald, theoretical physicist \\
      (classical general relativity, quantum field theory on curved
      spacetime)
    \end{flushright}
  \end{quote}
  \\
  \hline
  \begin{quote}
    It is tempting but conceptually problematic to think of black
    holes as objects in space, things that can move and be pushed
    around.  They are simply not quasi-localised lumps of any sort of
    `matter' that occupies [spacetime] `points'.
    \skipline[-.5]
    \begin{flushright}
      -- Domenico Giulini, theoretical physicist \\
      (classical general relativity, canonical quantum gravity,
      foundations of quantum mechanics)
    \end{flushright}
  \end{quote}
  \\
  \hline
  \begin{quote}
    One can try to define a black hole in the context of holography
    and AdS-CFT{} as a macroscopic $N$-body solution to the quantum
    field theory that evolves like a fluid on the boundary of
    spacetime, which one can argue are the only solutions with
    horizons in the interior.
    \skipline[-.5]
    \begin{flushright}
      -- Paul Chesler, theoretical physicist \\
      (numerical relativity, holography)
    \end{flushright}
  \end{quote}
  \\
  \hline
\end{tabular}

\noindent\begin{tabular}[c]{|p{\textwidth}|}
  \hline
  \begin{quote}
    In analog gravity things get more difficult, since the dispersion
    relation could mean that low energy waves cannot get out [of the
    horizon] while high energy ones can (or vice versa).
    \skipline[-.5]
    \begin{flushright}
      -- Bill Unruh, theoretical physicist \\
      (classical general relativity, quantum field theory on curved
      spacetime, analogue gravity)
    \end{flushright}
  \end{quote}
  \\
  \hline
  \begin{quote}
    The versions of the description [of black holes] used tacitly or
    explicitly in different areas of classical physics
    (\emph{e}.\emph{g}.\@ astrophysics and mathematical general
    relativity) differ in detail but are clearly referring to the same
    entities.
    \skipline[-.5]
    \begin{flushright}
      -- David Wallace, philosopher \\
      (foundations of quantum mechanics, statistical mechanics,
      cosmology)
    \end{flushright}
  \end{quote}
  \\
  \hline
\end{tabular}

\section*{Box 3: Quantum Gravity Views on Black Holes}

\begin{tabular}[c]{|p{\textwidth}|}
  \hline
  \begin{quote}
    I would not define a black hole [in this way]: by its classical
    central singularity.  To me it is clear that that is an artefact
    of the limitations of General Relativity, and including quantum
    effects makes it disappear.
    \skipline[-.5]
    \begin{flushright}
      -- Francesca Vidotto, theoretical physicist \\
      (loop quantum gravity, quantum gravity phenomenology)
    \end{flushright}
  \end{quote}
  \\
  \hline
  \begin{quote}
    A primary motivation of my research on quasi-local horizons was to
    find a way of describing black holes in a unified manner in
    various circumstances they arise in fundamental classical physics,
    numerical relativity, relativistic astrophysics and quantum
    gravity.
    \skipline[-.5]
    \begin{flushright}
      -- Abhay Ashtekar, theoretical physicist \\
      (classical general relativity, loop quantum gravity, cosmology)
    \end{flushright}
  \end{quote}
  \\
  \hline
  \begin{quote}
    Black holes are not clearly defined in string theory and
    holography.
    \skipline[-.5]
    \begin{flushright}
      -- Andy Strominger, theoretical physicist \\
      (string theory, holography)
    \end{flushright}
  \end{quote}
  \\
  \hline
  \begin{quote}
    [T]he event horizon \ldots \space is a
    \emph{spacetime concept}, and spacetime itself is a classical
    concept.  From canonical gravity we learn that the concept of
    spacetime corresponds to a particle trajectory in mechanics.  That
    is, after quantization the spacetime disappears in quantum gravity
    as much as the particle trajectory disappears in quantum
    mechanics.
    \skipline[-.5]
    \begin{flushright}
      -- Claus Kiefer, theoretical physicist \\
      (semi-classical gravity, canonical quantum gravity)
    \end{flushright}
  \end{quote}
  \\
  \hline
\end{tabular}

\newpage

\section*{Acknowledgements}

I am grateful to all the many physicists and philosophers who
responded to my questions with thoughtful enthusiasm---you are too
many to name, but you know who you are.  This essay would have been
much poorer without the illumination of your discussions.  I must,
however, single out Abhay Ashtekar, Beatrice Bonga, Paul Chesler, Bob
Geroch, Domenico Giulini, Gerard 't Hooft, Ted Jacobson, Claus Kiefer,
Avi Loeb, Juan Maldacena, Don Marolf, Ramesh Narayan, Daniele Oriti,
Carlo Rovelli, Karim Th\'ebault, Bill Unruh, Bob Wald, David Wallace,
and Shing-Tung Yau for supererogatory input and further discussion.  I
thank Bill Unruh and Bob Wald also for their recollections of the
attitude of relativists in the 1960s and 1970s to black holes, as well
as Avi Loeb and Ramesh Narayan for discussion about the reception of
the idea in the community of astrophysicists at the same time.  I am
also grateful to Marios Karouzos, Associate Editor at \emph{Nature
  Astronomy}, for suggesting I write this piece.

\noindent Some of this work was completed at the Black Hole Initiative
at Harvard University, which is funded through a grant from the John
Templeton Foundation.  The rest was completed at the Munich Center for
Mathematical Philosophy, in part funded by a grant from the Deutsche
Forschungsgemeinschaft (CU 338/1-1).

\sloppy

\end{document}